\documentclass{PoS}

\title{Performance of the SoLid Reactor Neutrino Detector}

\ShortTitle{Performance of the SoLid Reactor Neutrino Detector}

\author{\speaker{Luis Manzanillas}\thanks{On behalf of the SoLid  collaboration}\\
        LAL, Univ Paris-Sud, CNRS/IN2P3, Universit\'e Paris-Saclay, Orsay, France\\
        E-mail: \email{manzanillas@lal.in2p3.fr}}

\abstract{The SoLid  collaboration is currently operating a 1.6 tons neutrino detector near the Belgian BR2 reactor, 
with main goal the observation of the oscillation of electron antineutrinos to previously undetected 
flavor states. The highly segmented SoLid  detector employs a compound scintillation technology based on 
PVT scintillator in combination with a $^{6}$LiF:ZnS(Ag) screens containing $^{6}$Li isotopes. The experiment
has demonstrated a channel-to-channel response that can be controlled
to the level of a few percent, and energy resolution of better than 14\% at 1 MeV, and a 
determination of the interaction vertex with a precision of 5 cm.
In this contribution we will highlight the major outcomes of the R\&D program that preceded the
construction of the full-scale detector, the quality control during component manufacture and integration,
as well as the current performance and stability of the full-scale system. The possibilities for in-situ
calibration of the detector with various radioactive sources will be discussed as well.}

\FullConference{The 39th International Conference on High Energy Physics (ICHEP2018)\\
		4-11 July, 2018\\
		Seoul, Korea}

\begin{document}

\section{Introduction}
A broad experimental program is ongoing to search for light sterile $\nu$'s at very short baselines at nuclear reactors.
Recent evaluations of the experimental anomalies pointing to the existence of a light sterile $\nu$ state, suggest a $\Delta m^{2}_{41}$
of about 1.3 or 1.7 eV$^{2}$ and a small mixing angle \cite{Gariazzo:2017fdh,Dentler:2018sju}. 
It implies an oscillation length of the order of a few meters for $\nu$'s of about $3$ MeV energies, the 
typical value of $\overline{\nu}_e$'s emitted at nuclear reactors. 
Then a  good energy and spatial resolution are critical to avoid smearing this oscillatory behavior. 
A large reactor core produces large uncertainties in the distance traveled by the $\overline{\nu}_e$
from the reactor core to the detection point. Hence, the reactor core should be as compact as possible.
Only a few research reactor facilities allow to satisfy these conditions. The SoLid experiment is located at the 
Belgian BR2 nuclear plant. This site owns a very compact reactor core, whose effective diameter is smaller than 50 cm, 
allowing to place the SoLid detector at baselines between 6 and 9 meters. In addition, its fuel is highly enriched in $^{235}$U,
which reduces the contribution from other fissile isotopes whose $\overline{\nu}_e$ energy spectrum
is less well known than $^{235}$U \cite{Giunti:2017yid}. 
Conducting a $\nu$ experiment in surface, and very close to a reactor core, implies very high backgrounds induced by cosmic muons and 
by the reactor itself. Thus, this kind of experiment should provide as much tools as possible in order to discriminate the high backgrounds
in such conditions.

Reactor $\overline{\nu}_e$'s are detected via the Inverse Decay Process (IBD): $\overline{\nu}_e + p \rightarrow e^{+}+n$,
whose signature, are prompt - delayed signals correlated in time and space. 
The prompt signal being originated by the positron, while the delayed 
signal is produced a few $\mu$s later by the neutron capture after its thermalization.
In order to maximize the use of this information, the SoLid collaboration
has developed an innovative detection technique \cite{Abreu:2018pxg}.
\section{SoLid  technology}
\begin{figure}[h]
\centering
\begin{tabular}{c c}
  \includegraphics[width=0.3\linewidth]{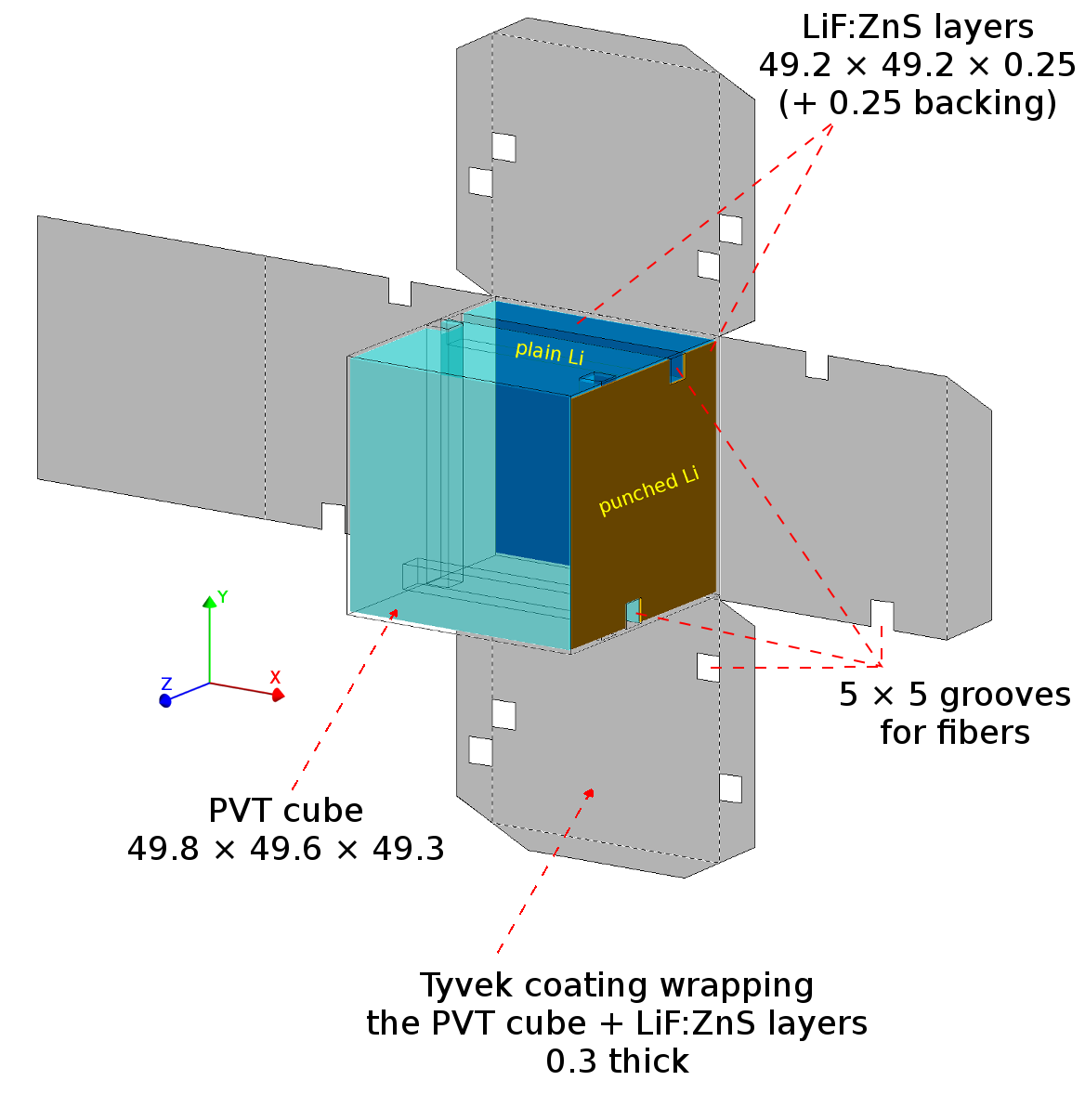} & \includegraphics[width=0.7\linewidth]{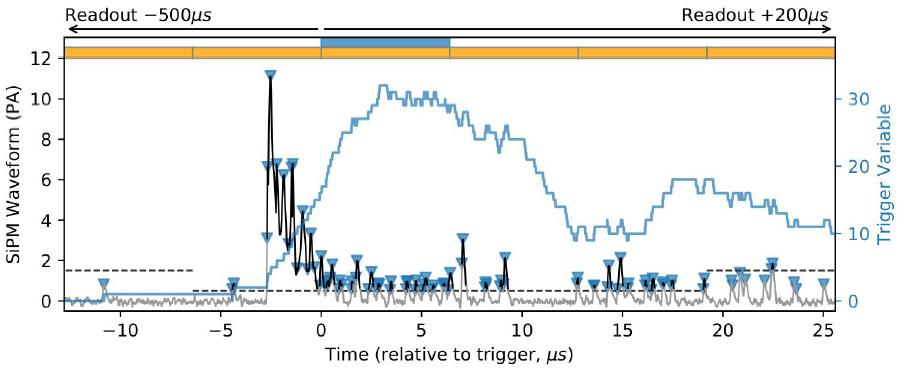}
\end{tabular}
\caption{Left : A SoLid cube. Right: Neutron trigger in SoLid.} 
\label{fig:nTrigger}
\end{figure}

In SoLid, $5\times5\times5$ cm$^{3}$ polyvinyltoluene (PVT) cubes are used as $\overline{\nu}_e$ target for the IBD reaction. PVT is 
used to detect and reconstruct the $e^{+}$ energy, providing the required information to assess the initial $\overline{\nu}_e$ energy.
PVT also serves as moderator for the neutron thermalization, which will be captured in a
second scintillator, giving rise to an unambiguous delayed signal. Thus PVT cubes are combined with two $^{6}$LiF:ZnS
screens  of $\sim$250 $\mu$m thickness, that are placed along two faces of each cube as shown in figure \ref{fig:nTrigger}. 
This ensemble is wrapped with Tyvek in order to guarantee an optical isolation among the cubes. Finally,
cubes are mounted in frames of 16$\times$16 cubes. Signals are read out by a network of wavelength shifting fibers,
consisting in a 2D array of 64 fibers per plane, which are coupled to an MPPC in one end and to 
a mirror in the other end \cite{Abreu:2018ajc}.
The SoLid detector consists of 50 planes, grouped into 5 modules, for a total mass of 1.6 tons.
The segmentation of the detector, and its hybrid technology,  provide powerful tools to discriminate backgrounds, but represents 
a real challenge in terms of construction, homogeneity, and calibration.

In the SoLid configuration, most of the neutrons are captured on the $^{6}$Li nuclei of the $^{6}$LiF:ZnS screens, because of its higher neutron capture cross
section. The $^{6}$Li nuclei then will break-up, depositing about 4.7 MeV in the ZnS scintillator.
The use of two different scintillators enables the unambiguous identification of prompt (PVT)
and delayed signals (ZnS). It is possible since the time profiles of the PVT and the ZnS scintillators are different,
$\mathcal{O}$(1) ns  and  $\mathcal{O}$(1) $\mu$s timescales respectively. In addition, 
the SoLid technology makes possible to set a dedicated neutron trigger, reducing considerably data rate. The neutron trigger consists in a simple algorithm that 
counts the number of peaks above a certain threshold for a given time window, as shown in figure \ref{fig:nTrigger} (right).
Prompt signals are recovered combining the neutron trigger with a large buffer, namely, 400 $\mu$s prior and 200 $\mu$s post neutron trigger are read out.
A high IBD efficiency is thus achieved.
\section{Quality assurance and commissioning}
\begin{figure}[h]
\centering
\includegraphics[width=\textwidth,trim={0cm 0.2cm 0cm 0.8cm},clip]{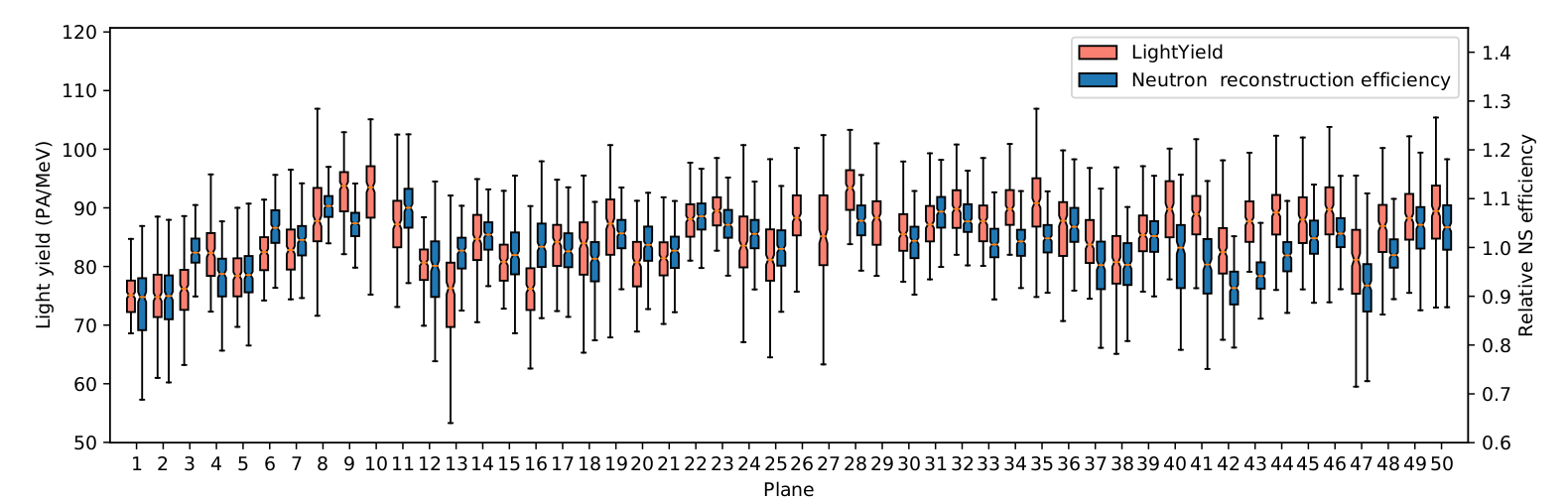}
\caption{Preliminary LY and relative neutron detection efficiency of the 50 planes of the SoLid detector.}
\label{CalibPrel}
\end{figure}

In order to attain a good homogeneity, and to guarantee the correct operation of all the components of the SoLid 
detector, a rigorous quality assurance process was developed during the detector construction \cite{Abreu:2018QA}.
To this end, an automated calibration system was designed and constructed. It allows to perform 
a preliminary calibration of each detector plane, by using a $^{22}$Na and AmBe/$^{252}$Cf radioactive sources, for a 
gamma and neutron calibration respectively. The goal of this calibration being the validation of the SoLid minimal physics requirements, and the identification of possible defective components.
Figure \ref{CalibPrel} shows the preliminary results of light yield (LY) and relative neutron detection efficiency of the 50 SoLid planes. This calibration has shown that 
a good homogeneity has been achieved. In addition, a LY and a neutron detection efficiency better than 60 PA/MeV  and 60 \% respectively have been reached.

During the quality assurance process, some minor problems where identified and fixed. Particularly important was the 
identification of a defective batch of Li screens. This problem was identified during the neutron calibration. Some cubes
were showing a very low neutron detection efficiency compared with the rest of the cubes, while the LY was showing a 
normal behavior. More detailed analysis revealed that these cubes were mounted with screens that had been half doped with $^{6}$Li. All the cubes 
mounted with these problematic screens were identified and replaced with new ones.
All the planes underwent the quality assurance process, and the first modules were mounted, deployed and commissioned at BR2 in November of 2017.
However, the commissioning of the full detector was completed in February of 2018. By March 2018, a first in-situ calibration was performed using 
$^{22}$Na, $^{252}$Cf, and AmBe radioactive sources. This calibration showed a LY and a neutron detection efficiency larger than 70 PA/MeV
and 75\% respectively. In addition, a very good agreement was found between the 1.27 MeV ($^{22}$Na) and the 4.4 MeV (AmBe) $\gamma$'s, confirming a linear energy 
response of the PVT. Further calibrations are expected to take place in 2018, with a complete set of $\gamma$ sources in order to test the energy response in a wider range.
Thus, $^{137}$Cs, $^{22}$Na, $^{60}$Co, $^{207}$Bi, AmBe, and $^{252}$Cf will be used. In addition, thanks to the segmentation of the detector,
muon tracks can be accurately reconstructed. This information
can be used to monitor the detector response in a daily basis, and to study the detector response at high energy.
\section{Conclusions}
The SoLid collaboration has developed a new detector concept for reactor $\overline{\nu}_e$ detection.
It employs a highly segmented hybrid technology.
Using a rigorous quality assurance process, the SoLid  collaboration has shown that 
the construction of a detector with a highly
segmentation, and with a homogeneous response can be successfully conducted. The construction and commissioning of
the SoLid Phase~1 (1.6 tons) has been completed in February of 2018. 
A first in-situ calibration revealed a very good performance, with a LY and a neutron detection efficiency
larger than 70 PA/MeV and 75 \% respectively. The detector is taking data in physics mode with a very good stability; $\sim$150 
days of reactor ON data are expected by the end of 2018, which will be used for a first physics result at the beginning of 2019.

\end{document}